\documentclass[aps,superscriptaddress,twocolumn,preprintnumbers,floatfix,showpacs,prx]{revtex4-1}

\usepackage{amsmath,amssymb}
\usepackage{dcolumn}
\usepackage{braket}
\usepackage{dsfont}
\usepackage{feynmp}
\usepackage{hhline}
\usepackage{graphicx}
\usepackage{mathtools}
\usepackage{xfrac}
\usepackage{paralist}
\usepackage{enumitem}
\usepackage[normalem]{ulem}
\usepackage{color}
\definecolor{darkblue}{rgb}{0.,0.,0.4}
\definecolor{darkred}{rgb}{0.5,0.,0.}

\newcommand{\beq}{\begin{eqnarray}}
\newcommand{\eeq}{\end{eqnarray}}

\begin{document}

\title{Cryogenic Cooling and Power Generation Using Quantum Hall Systems}
\author{Liang Fu}
\affiliation{Department of Physics, Massachusetts Institute of Technology, Cambridge, MA 02139, USA}
%\date{\today}
\begin{abstract}
The possibility of using quantum Hall systems for thermoelectric energy conversion is investigated. It is shown that the massive degeneracy and the metallicity of a partially-filled Landau level enable thermoelectric cooling and power generation with unprecedented efficiency at low temperature. The figure of merit is explicitly derived for a transverse thermoelectric device using the $\nu=0$ quantum Hall state of Dirac materials at charge neutrality, where due to electron-hole symmetry electrical Hall effect vanishes but thermoelectric Hall effect peaks.
\end{abstract}
\maketitle

Solid-state thermoelectric cooling is based on the generation of a heat current by passing an electrical current through a material, or the Peltier effect. A related technology is thermoelectric power generation that is based on the reverse phenomenon that a voltage difference appears between two sides of a material in proportion to their temperature difference, or the Seebeck effect.
The main advantages of thermoelectric coolers and generators are lack of moving parts, small size and stability. However, their efficiency is low because
both Peltier and Seebeck effects arise from the mutual coupling between heat and electrical currents, which is weak in most solids.
The search for highly efficient thermoelectric materials is a subject of vast research \cite{Goldsmid}.

The rapid advance of quantum (nano)electronics, infrared detection, superconducting qubit technology and space applications relies on low temperature refrigeration. However, thermoelectric cooling at low temperature is challenging due to the reduction of thermally excited charge carriers. The lowest temperature that can be reached with commercial 6-stage Peltier coolers is 170K.

In this work, we propose a method of thermoelectric cooling and power generation using Landau levels under a quantizing magnetic field. The advantage is twofold.
A partially filled Landau level in two-dimensional (2D) systems exhibits massive ground state degeneracy in clean noninteracting limit. Provided that disorder and electron interaction are weak, the entropy per charge carrier remains finite down to very low temperature. %This leads to a Nernst signal $S_{xy}$ and/or Seebeck coefficient $S_{xx}$ that is $T$-independent as opposed to $T$-linear.
%Moreover, a partially-filled Landau level system is an electrical conductor.
We show that because of its non-vanishing entropy and metallicity, a partially filled Landau level enables thermoelectric cooling and power generation with unprecedented efficiency at low temperature.

The use of magnetic fields to improve thermoelectric efficiency has a long history \cite{Behnia}.
Continuous cooling from room temperature to around $100$K was demonstrated decades ago using the giant Nernst effect in bismuth-antimony alloy under a modest field  \cite{cooling-bismuth}. A recent theoretical work proposed thermoelectric applications using the large, non-saturating Seebeck effect of Dirac/Weyl semimetals in the high-field quantum limit when electrical resistivity is dominantly transverse \cite{Skinner}.
These previous works consider three-dimensional (3D) materials with a {\it finite} density of states at the Fermi level. This sets a fundamental limit that  Seebeck and Nernst response are proportional to $T$, making the figure of merit $zT \propto T^2$ degrade rapidly as $T\rightarrow 0$ \cite{bismuth-zt}. In this work, we show that Landau levels in two dimensions offers an ideal solution to this challenge by providing a flat band with a singularly large density of states, while maintaining the metallicity.

We begin with some general consideration of coupled electrical and heat transport under a magnetic field $\bf B$. Throughout this work, we consider a quantizing field---or equivalently weak disorder---that satisfies $\mu B \gg  1$, where $\mu$ is the carrier mobility. Under this condition, Landau levels are formed and separated by cyclotron energy $\hbar \omega_c \gg \Gamma$, where $\Gamma$ is disorder-induced Landau level broadening.
%We will study and compare thermoelectricity at temperatures both larger and smaller than cyclotron energy.

The coupling between electricity and heat is often described in terms of  Seebeck and Nernst signals $S_{ij}$, which measure the voltage generated by a temperature difference under open-circuit condition. We find it more convenient to think in terms of thermoelectric conductivity $\alpha_{ij}$, which measures the electrical current $\bf I$ generated by a temperature gradient $\nabla T$ in the absence of any voltage (short-circuit condition): $I_i = -\alpha_{ij} \nabla_j T$. $S_{ij}$ and $\alpha_{ij}$ are related to each other: $S_{ij}= \alpha_{ik} \rho_{kj}$ where $\rho$ is resistivity. $S_{ij}$ and $\alpha_{ij}$ are related to each other: $S_{ij}= \alpha_{ik} \rho_{kj}$ where $\rho$ is resistivity.

In the presence of a magnetic field, $\alpha_{ij}$ generally has a part that is odd in $\bf B$, which we refer to as thermoelectric Hall conductivity. For an isotropic system under a field parallel to the $z$ axis, it appears as the off-diagonal $xy$ component satisfying  $\alpha_{xy}( B)=-\alpha_{yx}(B)=-\alpha_{xy}(-B)$.
By Onsager relation, $\alpha$ also measures the heat current $\bf Q$ generated by an electric field under isothermal condition: $Q_i= T \alpha_{ji}(-B)  E_j=T \alpha_{ij}(B)  E_j$. Since heat current is carried by thermal excitations, thermoelectric conductivity---both $\alpha_{xx}$ and $\alpha_{xy}$---is a Fermi surface property, in contrast with Hall conductivity and Seebeck/Nernst response.

\begin{figure*}%[!b] \centering
\includegraphics[width=0.9\textwidth] {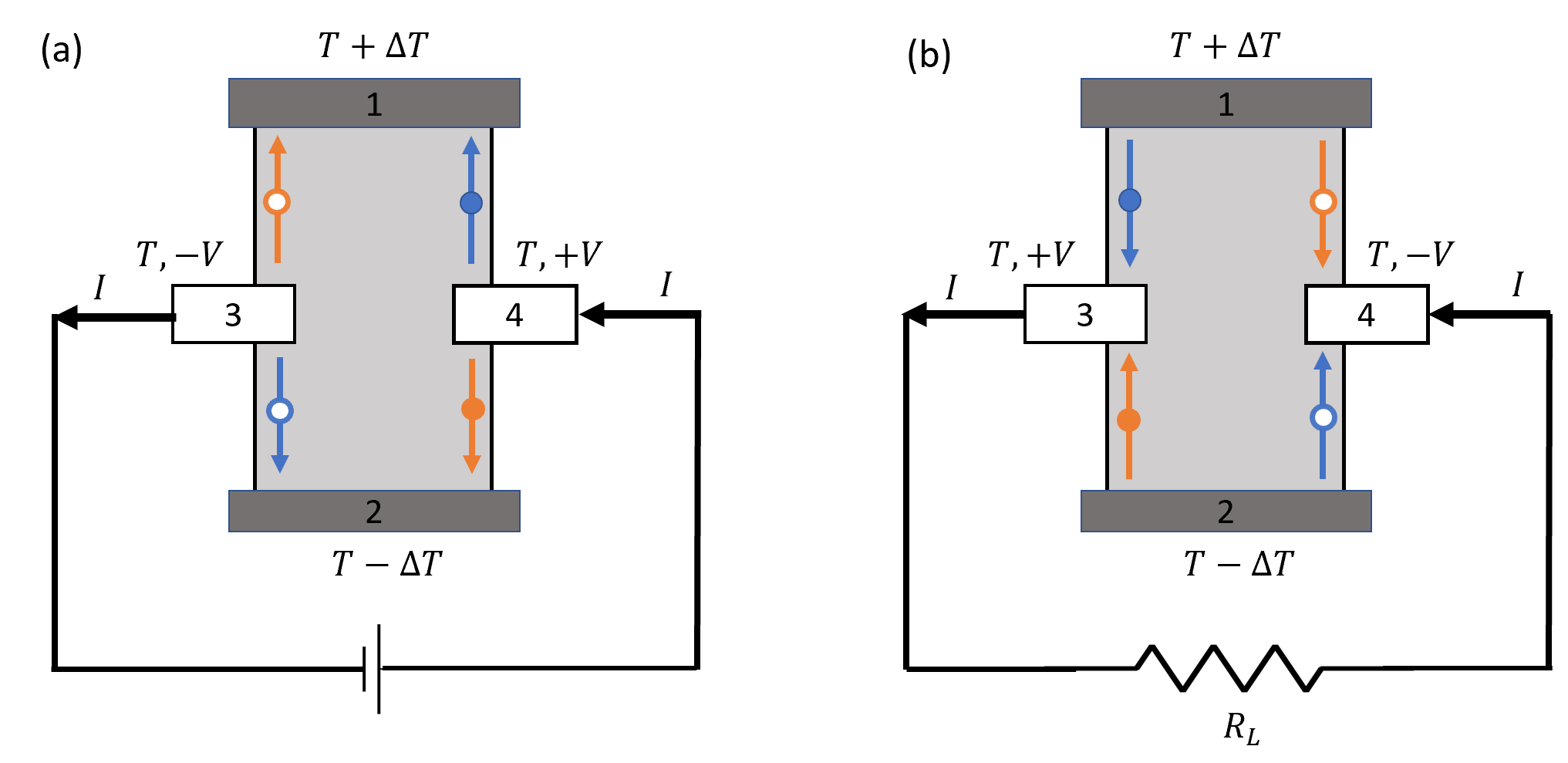}
\caption{Thermoelectric cooler (a) and power generator (b) based on a $\nu=0$ quantum Hall state having ambipolar edge modes with opposite chirality at energy $E>0$ and $E<0$. In the working state of the cooler or generator, transport current arises from excess charge $e$ and $-e$ carriers (denoted by dots and circles respectively) produced by temperature gradient and voltage. Note a decrease in the occupation of $E<0$ modes amounts to an increase of charge $e$ carrier relative to thermal equilibrium. }
\end{figure*}

At temperatures $k_B T \gg \hbar \omega_c$,  Landau level density of states  are thermally smeared. In this semiclassical regime, under an applied electric field ${\bf E}$ perpendicular to $\bf B$, charge carriers acquire a drift velocity ${\bf v}_d$ perpendicular to both field directions, which in the clean limit is solely determined by the balance of electric force and Lorenz force: ${\bf v}_d = {\bf E \times B}/B^2$.
This creates both a transverse electrical current and a transverse heat current. The latter is given by ${\bf Q}=T s {\bf v}_d$, where $s$ is the entropy density. Therefore, under a quantizing magnetic field, thermoelectric Hall conductivity is given by
\beq
%\sigma_{xy} &\equiv &   \frac{n_q}{B}, \label{sigma} \\
\alpha_{xy} &=&  \frac{s}{B}. \label{hall}
\eeq
Since entropy is associated with the number of thermal excitations within the energy $k_B T$ from the Fermi level, in the temperature range $\hbar \omega_c \ll k_B T \ll E_F  $ ($E_F$ is Fermi energy), we have $\alpha_{xy}\propto s \propto k_B T$ for metals and degenerate semiconductors.
%Therefore in this temperature range, $\alpha_{xy}$ decreases linearly with temperature. %On the other hand, electrical resistivity

In the dissipationless limit $\mu B \rightarrow \infty$ or $\Gamma/\hbar \omega_c  \rightarrow 0$, the thermodynamic formula for $\alpha_{xy}$ (\ref{hall}) continues to hold at low temperature $k_B T \ll \hbar \omega_c$ \cite{Girvin,Halperin, Kozii}. In this case, however, the entropy is strongly modified by Landau quantization of density of states. In two dimensional systems, the density of states is a sum of delta-shaped peaks centered at a discrete set of energies. At $k_B T \gg \Gamma$, when the Fermi energy is at the center of a Landau level, each Landau orbital has probability $1/2$ of being occupied and of being empty, resulting in a maximum entropy density $s=(\log 2) k_B (B/\Phi_0)$ where $\Phi_0=h/e$ is flux quantum. Therefore, in the temperature range $\Gamma \ll k_B T \ll \hbar \omega_c$, thermoelectric Hall conductivity of a quantum Hall system is peaked whenever the Landau level at Fermi energy is half-filled, and the peak value is universal
\beq
\alpha_{xy} =  \frac{(\log 2) k_B e}{h} \label{alpha}
\eeq

%where $g_L$ is an integer accounting for possible spin/valley degeneracy. %Meanwhile, $\alpha_{xx}=0$ due to the emergent particle-hole symmetry.
The result (\ref{alpha}) was implicitly contained in the seminal work of Girvin and Johnson \cite{Girvin}, although it focuses on the Seebeck coefficient $S_{xx}$. The latter depends on the total Landau level filling and hence is not as universal as $\alpha_{xy}$. For example, as we will see later, in the quantum Hall regime of graphene at charge neutrality, $S_{xx}=0$ while $\alpha_{xy}$ remains given by (\ref{alpha}).

Thanks to the finite $\alpha_{xy}$, 2D quantum Hall systems are advantageous for thermoelectric energy conversion at low temperature over traditional thermoelectric materials that employ $\alpha_{xx}$. The latter decreases linearly with temperature when $k_B T \ll E_F$,  as seen from the generalized Mott formula $\alpha_{xx}= (-\pi^2 k_B^2 T/3e) d \sigma(E)/dE |_{E_F} $ where $\sigma(E)$ is energy-dependent conductivity. For 3D systems under a magnetic field, the continuous energy spectrum of one-dimensional Landau band dispersing along the field direction leads to $\alpha_{xy} \propto T$ \cite{Vadim, Kozii} which decreases at low temperature, in contrast with the 2D case.

Motivated by the consideration of $\alpha_{xy}$, we propose a thermoelectric cooler and a power generator based on quantum Hall systems.
%This work we choose a Nernst device geometry as shown in Fig.1.
As shown in Fig.1, a quantum Hall system is contact with two heat baths 1 and 2 having different temperatures. Each bath exchanges energy the system without transferring any net charge. Our system is also connected via electrical leads 3 and 4 to an external circuit, a battery in the case of a cooler (Fig.1a) and a resistance load in the case of a generator (Fig.1b).

Power generation is achieved by natural heat flux from hot to cold baths, which produces a voltage between the leads and thus supplies electrical power to the external resistor, see Fig.1b. On the other hand, passing a sufficiently large electrical current between the leads cools the cold bath by directing heat from it into the hot bath against the opposing temperature difference, see Fig.1a.
In our device, electrical and heat currents run in orthogonal directions. For such transverse thermoelectric geometry, there is no need of employing both n and p-type materials as traditional Peltier coolers and Seebeck generators do.

In thermal equilibrium all terminals are at the same temperature  $T_j = T$ and chemical potential $\mu_j=\mu$, $j=1,...,4$. In the working state of our device, $T_j$ and $\mu_j$ will generally be different from their equilibrium values and we write $T_j = T + \Delta T_j, \mu_j = \mu+ e V_j$. $\Delta T_j$ and $V_j$ are called generalized forces. There will also be net charge and heat currents, denoted as $I_j$ and $Q_j$, that flow within the quantum Hall system into (defined as positive) or out of (defined as negative) the terminals. When the deviation from equilibrium is small, the currents and forces are linearly related to each other by transport coefficients of our multi-terminal quantum Hall system.

For simplicity we assume our system has a twofold rotation symmetry that exchanges terminals $1\leftrightarrow 2$ and $3\leftrightarrow 4$ at opposite ends.
Then, in the working state of our device, the currents and forces at terminal 1 (3) are opposite to those at 2 (4), which we write as
$
J_{1} = -J_2 \equiv -J_y,
J_{4} = - J_3 \equiv -J_x,
$
and
$
F_{1} = - F_2 \equiv F_y/2,
F_{4} = - F_3 \equiv  F_x/2,
$
where $J$ stands for $I$ or $Q$, and $F$ for $V$ or $\Delta T$.
By definition, there is no net charge current flowing into a heat bath, so $I_y=0$.
We also assume that the two electrical leads are at the same temperature (as is the case when the external circuit is a perfect thermal conductor), so $\Delta T_x =0$.

%The currents and forces are linearly related to each other by
%\beq
%I_i &=&  \sum_{j} G_{ij} V_j + L^{eh}_{ij} \Delta T_{j} \nonumber \\
%Q_i &=&  \sum_{j} T L^{he}_{ij} V_j + K_{ij} \Delta T_{j} \label{transport}
%\eeq
%
For a given temperature difference $\Delta T_y$ between hot and cold baths, solving the coupled electrical/thermal transport equation under the condition $I_y=0$ will yield the electrical current $I_x$ and the heat current $Q_y$ in the cooler for a given voltage $V_x$ set by external battery.  Likewise, for a generator with an external resistance $R_L$,  one can obtain $I_x$ and $Q_y$ by further using Ohm's law $I_x = V_x/R_L$. % then allow us to evaluate the thermoelectric efficiency of our device.

To simplify our analysis further, below we consider the quantum Hall regime of electron-hole balanced systems. Particularly suitable for our purpose are two dimensional systems with massless Dirac dispersion, such as graphene and topological insulator thin films including HgTe \cite{HgTe} and Bi$_2$Se$_3$ \cite{Ando}. Under a magnetic field, these Dirac materials exhibit a special $n=0$ Landau level at zero energy, which is exactly half-filled at charge neutrality and lends itself equally to electron and hole excitations.

On general ground, electrical Hall conductivity, thermal Hall conductivity (denoted by $\kappa_{xy}$) as well as diagonal thermoelectric conductivity and Seebeck coefficient are odd under charge conjugation.  In contrast, thermoelectric Hall conductivity is invariant under charge conjugation. As a result, the $\nu=0$ quantum Hall state at charge neutrality has $\sigma_{xy}=\kappa_{xy}=0$  and $\alpha_{xx}=S_{xx}=0$ due to electron-hole cancellation, but a nonzero $\alpha_{xy}$ that takes the universal value under the specified condition. Thus, thermoelectric Hall effect is the only Hall response remaining at $\nu=0$!

%Before proceeding, we call attention to a distinctive feature of compensated semimetals, that is, the coexistence of electrons and holes has opposite effects on electrical and thermoelectric transport. Under crossed electric and magnetic fields, electrons and holes drift in the same transverse direction $\bf E \times \bf B$, thus producing electrical current in opposite directions and heat (=entropy) current in the same direction \cite{Skinner}. As a result, charge Hall effect is suppressed by their cancellation, while thermoelectric Hall effect is enhanced by their addition \cite{Kozii}. We note in passing that the simultaneous occurrence of $\sigma_{xx} \gg \sigma_{xy}$  and $S_{xy} \gg S_{xx}$ resulting from electron-hole balance was indeed observed in bulk bismuth \cite{bismuth}, graphite \cite{Behnia} and WTe$_2$ \cite{Felser} under modest magnetic fields.

For such electron-hole-balanced system, the simplified transport equation with $\sigma_{xy}= \kappa_{xy} = \alpha_{xx}=0$ takes the general form
\beq
I_x &=& G V_x + L^{eh}  \Delta T_{y} \nonumber \\
Q_y &=&  -T L^{he}  V_x + \tilde{K}  \Delta T_{y}. \label{IQ}
\eeq
where $G$ is longitudinal electrical conductance, $\tilde{K}$ longitudinal thermal conductance in the absence of any voltage, and $L^{eh}, L^{he}$ are transverse thermoelectric conductance.

To understand how our device enables thermoelectric cooling and power generation, it is useful to consider two limits first. In the presence of a temperature difference $\Delta T_y$, when we set $R_L=0$ to short-circuit the generator,  a transverse electrical current is produced by thermoelectric Hall effect, denoted as $I_x^0$. Alternatively, in an open circuit with $R_L=\infty$, an {\it opposing} voltage difference $V_x^0$ arises to cancel the electrical current that would otherwise be present and enforces $I_x=0$. For finite $R_L$,  $I_x$ will be nonzero and smaller than $I_x^0$. The ratio of output electrical power and the heat current $|Q_y|$ defines the coefficient of performance,
$
%\beq
\phi_p = I_x^2 R_L/|Q_y|.
%\eeq
$
$\phi_p$ will be maximum at a certain load resistance.

In the cooling mode, the external battery sets a forward voltage $V_x$ (opposite to $V_x^0$) to increase the electrical current $I_x$ above the short-circuit current $I_x^0$. When $V_x$ is sufficiently large, thermoelectric Hall effect generates a heat current $Q_y$ going from cold to hot bath against the opposing temperature difference. The cooling power $q$ is given by $Q_y$ minus Joule heating in the cold bath, $q=Q_y - I_x V_x/2$ (note that half of $I_x$ goes through the cold bath).
The coefﬁcient of performance  is defined as the ratio of $q$ and the rate of electrical power input,
$
%\beq
\phi_c = (Q_y - I_x V_x/2)/(I_x V_x).
%\eeq
$
Since $Q_y \propto V_x$ and Joule heating $\propto V_x^2$, $\phi_c$ will be maximum at a certain applied voltage $V_x$.

Following standard analysis \cite{Goldsmid}, it can be shown that the maximum coefficient of performance $\phi_c$ or $\phi_p$ of our cooler or generator increases with a dimensionless quantity $ZT$ known as the figure of merit,
%find the optimum coefficient of performance \cite{Goldsmid}
%\beq
%\phi_p^{max} &=& \frac{\Delta T}{T_1} \frac{\sqrt{1+ZT}-1}{\sqrt{1+ZT}+T_2/T_1}, \nonumber \\
%\phi_c^{max} &=& \frac{T_1}{\Delta T} \frac{\sqrt{1+ZT}-T_2/T_1}{\sqrt{1+ZT}+1}
%\eeq
%with $T_{1,2} = T+ \Delta T/2$. Here, $ZT$ is the dimensionless figure of merit defined as
\beq
ZT = \frac{L^{eh}  L^{he}  T}{G K}
\eeq
where $T = (T_1 + T_2)/2$ is the mean temperature of heat baths; $K= \tilde{K} + T L^{eh} L^{he}/G $ is thermal conductance in the absence of any electrical current, i.e., in ``open circuit'' condition. For a cooler, the maximum temperature difference attainable is $\Delta T_{max} = Z T_1^2/2$. For a generator, the efficiency approaches the Carnot limit of $(T_1 - T_2)/T_1$ as $ZT \rightarrow \infty$.

$ZT$ depends on a combination of electrical, thermal and thermoelectric conductances of the quantum Hall system in the four-terminal geometry shown in Fig.1.
If the system size is large enough,  the  conductance is proportional to the local conductivity. Importantly, half-filled Landau level at charge neutrality shows a finite, $T$-independent longitudinal conductivity $\sigma$ at low temperature, which is experimentally found to be on the order of $e^2/h$ in graphene and topological insulator thin films. Assuming lattice thermal conductivity is negligible at low temperature, it follows from Wiedemann-Franz law that $\kappa$ is on the order of $(k_B/e)^2 T \sigma$. Using these values for $\sigma$ and $\kappa$ along with the universal value of $\alpha_{xy}$ given in (\ref{alpha}), we conclude that thermoelectric figure of merit $ZT$ is of order unity throughout the temperature range $\Gamma \ll k_B T \ll \hbar \omega_c$.

We further study $ZT$ in phase-coherent transport regime where electrical, thermal and thermoelectric conductances are all determined by transmission probability of an electron or a hole going from one terminal to another. In the weak disorder limit considered here, conductance is dominated by edge-state transport. This allows us to calculate $G, K, L^{eh}, L^{he}$ explicitly using scattering approach \cite{scattering} and thus determine $ZT$.

The zero-energy $n=0$ Landau level in graphene, HgTe quantum well and topological insulator thin film all have a twofold degeneracy, which is associated with
valley, spin and top/bottom surface layer degrees of freedom respectively. This degeneracy is split at the edge of the sample, giving rise to two branches of edge modes, counter-clockwise at $E>0$ and clockwise at $E<0$. As a result, electrons and holes go in opposite directions at the edge, in tandem with the sign change of quantized Hall conductance. We emphasize that the existence of these ``ambipolar'' edge states within the energy gap between $n=0$ and $n=\pm 1$ Landau levels is guaranteed by topology: it is required by the first quantized Hall plateau on the electron and hole side. At a given energy, the edge state is chiral and elastic backscattering from impurity is forbidden.
%When disorder-induced mixing of the two degenerate $n=0$ Landau level components is negligible, edge states exist at all energies $E \neq 0$.

Due to its chirality, the transmission probability of an electron that occupies an $E>0$ edge mode going from terminal $i$ to $j$ is $1$ if $j$ is a downstream neighbor of $i$, and $0$ otherwise. In contrast, due to its opposite chirality, the transmission probability of a hole---the state of having an unoccupied $E<0$ edge mode---going from terminal $i$ to $j$ is $1$ if $j$ is an upstream neighbor of $i$, and $0$ otherwise.

Therefore, applying a voltage or temperature change at a given terminal can only produce nonzero electrical and/or heat currents at its downstream neighbor, at its upstream neighbor, and at itself---the last being a sum of the first two by the law of current conservation.
The downstream (upstream) current is solely carried by electron (hole) and thus depends only on the change of occupation of $E>0$ ($E<0$) edge modes due to $\Delta \mu = e V$ or $\Delta T$.

To obtain electrical conductance $G$, we calculate the electrical current $I_4$ produced by $V_4$. For $e V_4>0$ as shown in Fig.1a, the increase of chemical potential at lead 4 sends more electrons to bath 1 and less holes to bath 2, which add up to yield the electrical current:
\beq
 I_4 &=&   \int_{0}^\infty  d E   (\frac{e}{h}) (- \frac{\partial{f}}{\partial E} eV_4) +  \int_{-\infty}^{0}   d E  (\frac{-e}{h}) (\frac{\partial{f}}{\partial E} e V_4)  \nonumber
\eeq
where $f(E)=1/(\exp(\beta E)+1)$ is the Fermi-Dirac distribution at charge neutrality.
Note that the change of electron occupation is $\delta f $, while the change of hole occupation is $-\delta f$.  %Here we assume the $n=0$ Landau level is twofold degenerate, i.e., $g_L=2$.
From this current-voltage relation we obtain
\beq
G = I_4/(2V_4) = (1/2) e^2/h
\eeq
The prefactor $1/2$ is due to two resistors in series in the four-terminal geometry.

Similarly,  we calculate the heat current $Q_1$ produced by the temperature change $\Delta T_1$. As shown in Fig.1b, the increase of temperature at bath 1 sends more electrons from lead 3 and more holes to lead 4, which add up to yield the heat current
\beq
Q_1
=     \int_{0}^\infty  d E   (\frac{E}{h}) (- \frac{\partial{f}}{\partial E} \frac{E \Delta T_1}{T})      +  \int_{-\infty}^{0}   d E  (\frac{-E}{h}) (\frac{\partial{f}}{\partial E} \frac{E \Delta T_1}{T}) \nonumber
\eeq
where we have used identity $ \partial f/\partial T = -  \partial f/\partial E \cdot  (E/ T)$. Note that a hole that corresponds to an unoccupied $E<0$ mode is an excitation that costs energy $-E>0$. The thermal conductance is then given by
\beq
\tilde{K} = \frac{1}{2} \frac{\pi^2}{3} \frac{k_B^2 T}{h}
\eeq

Finally, we calculate the  electrical current $I_3$  produced by opposite temperature changes at the two baths: $\Delta T_1 =  - \Delta T_2 \equiv  \Delta T$.
As shown in Fig.1b, the increase of temperature at bath 1 sends more electrons to lead 3, while the decrease of temperature at bath 2 sends less holes to lead 3. The two contributions  add up to yield the electrical current:
\beq
I_3 =  \int_{0}^\infty  d E   (\frac{e}{h}) (- \frac{\partial{f}}{\partial E} \frac{E \Delta T}{2T})  +  \int_{-\infty}^{0}   d E  (\frac{-e}{h}) (\frac{\partial{f}}{\partial E }  \frac{E (-\Delta T)}{2T} ).  \nonumber
\eeq
From this we obtain the thermoelectric Hall conductance
\beq
L^{eh} =  \log(2) k_B e/h.
\eeq
Likewise, from the heat current $Q_3$ produced by concurrent voltages $V_1 = -V_2$, we find $L^{he}=L^{eh}$, in accordance with the general symmetry property in scattering theory of thermoelectric transport \cite{Butcher}.

Putting these conductance values together, we find $ZT$ is a temperature-independent constant:
\beq
ZT = \frac{\log^2(2)}{\frac{1}{2} (\frac{\pi^2}{6} + 2 \log^2(2))} \approx 0.37.
\eeq
This result is independent of additional degeneracy of the $n=0$ Landau level that may be present (e.g., spin degeneracy in addition to the aforementioned valley degeneracy in graphene), because such degeneracy increases electrical conductance, thermal conductance and thermoelectric Hall conductance by the same factor without affecting $ZT$. Since a large electronic thermal conductance is desirable in order to outweigh the phonon contribution, a large Landau level degeneracy is advantageous.

We now discuss the conditions for achieving such a record-high thermoelectric figure of merit with existing materials. Graphene, HgTe and Bi$_2$Se$_3$ thin films all have a high Dirac velocity on the order of $10^6$m/s. A $1$T magnetic field already creates an energy gap about $400$K between $n=0$ and $n=\pm 1$ Landau level. In high-mobility samples, the Landau level width is about $10$K \cite{width1, width2, width3}. Thus, we expect thermoelectric cooling and power generation to be efficient  in a wide range of temperature $\Gamma <k_B T < \hbar \omega_c$.

It is encouraging that ample evidence of electrical transport mediated by ambipolar edge states has been observed in the $\nu=0$ quantum Hall state in graphene \cite{scape, nonlocal}, HgTe \cite{HgTe1, HgTe2, HgTe3} and Bi$_2$Se$_3$ thin films \cite{Chen}. Moreover, a peak of $\alpha_{xy}$  approaching the universal value (\ref{alpha}) was observed at $\nu=0$ in graphene  \cite{Ong} (see also \cite{Kim,Peng}) and bilayer graphene \cite{bilayer} in the temperature range specified above.

In practice, we envision using 2D quantum Hall systems for thermoelectric cooling of small quantum devices. In order to cool a 3D bath at low temperature, it may be useful to employ bulk crystals formed with weakly coupled layers, such as  graphite \cite{Geim}, ZrTe$_5$ \cite{Zhang} and organic molecular crystals \cite{organic},  where 3D quantum Hall states have recently been observed.

While our detailed analysis focused on electron-hole-symmetric $\nu=0$ quantum Hall state, our conclusion that $ZT$ remains finite at low temperature follows from two essential features of all partially filled Landau levels (more generally flat bands with Chern number): (1) a finite $\alpha_{xy}$ due to massive degeneracy and (2) a finite electrical conductivity $\sigma_{xx}$ due to its metallicity, together with Wiedemann-Franz law.

Last but not the least, throughout this work we have neglected the role of Coulomb interaction in lifting Landau level degeneracy.  The characteristic energy scale associated with Coulomb interaction is $e^2/\epsilon l_B$, where $\epsilon$ is dielectric constant and $l_B$ is magnetic length. For the purpose of thermoelectric cooling and power generation,  this energy scale can be sufficiently suppressed by strong dielectric screening or by working with a small magnetic field. On the other hand, the physics of thermoelectric response of interacting electrons in partially-filled Landau level is a fascinating topic that deserves more study in future.

\textit{Acknowledgments}
I thank Patrick Lee for an insightful discussion, Duncan Haldane for his interest, Vlad Kozii, Brian Skinner, Liyuan Zhang, Xiaosong Wu and Donna Sheng for collaborations on related projects, and Gang Chen for initiating my interest in thermoelectricity. I thank the hospitality of Institute of Condensed Matter Theory at University of Illinois Urbana-Champaign and Princeton Center of Theoretical Science, where this work was conceived.

This work was supported by DOE Office of Basic Energy Sciences, Division of Materials Sciences and Engineering under Award $\text{DE-SC0018945}$. LF was supported in part by the David and Lucile Packard Foundation.

\end{document}